\documentclass[%
preprint,
showpacs,preprintnumbers,
nofootinbib,
 amsmath,amssymb,
 aps,
]{revtex4-1}
\usepackage{graphicx}% Include figure files
\usepackage{dcolumn}% Align table columns on decimal point
\usepackage{bm}% bold math
\usepackage{url,hyperref}

\begin{document}

% Use the \preprint command to place your local institutional report
% number in the upper righthand corner of the title page in preprint mode.
% Multiple \preprint commands are allowed.
% Use the 'preprintnumbers' class option to override journal defaults
% to display numbers if necessary
\preprint{IPMU13-0153, IISER-K-08-13, OSU-HEP-13-05}
\newcommand{\newc}{\newcommand}
\def\lum             {{\cal L}}
\newc{\ra}{\rightarrow}
\newc{\gl}   {\mbox{$\wt{g}$}}
\newc{\mgl}  {\mbox{$m_{\gl}$}}
\def \met  {\mbox{${E\!\!\!\!/_T}$}}
\newc{\wt}{\widetilde}
\def \mlspj{m_{\lspj}}
\def \lspone{\wt\chi_1^0}
\def \mlspone{m_{\lspone}}
\newc{\ifb}{\mbox{${\rm fb}^{-1}$}}
\def \etslash{\not \! E_T }
\newc{\del}{\delta}

%Title of paper
\title{Compressed SUSY at 14 TeV LHC}

% repeat the \author .. \affiliation  etc. as needed
% \email, \thanks, \homepage, \altaffiliation all apply to the current
% author. Explanatory text should go in the []'s, actual e-mail
% address or url should go in the {}'s for \email and \homepage.
% Please use the appropriate macro foreach each type of information

% \affiliation command applies to all authors since the last
% \affiliation command. The \affiliation command should follow the
% other information
% \affiliation can be followed by \email, \homepage, \thanks as well.
\author{Biplob Bhattacherjee}
\affiliation{Kavli IPMU (WPI), The University of Tokyo, Kashiwa, Chiba 277-8583, Japan}
\email{biplob.bhattacherjee@ipmu.jp}

\author{ Arghya Choudhury }
\affiliation {Department of Physical Sciences,\\ 
	  Indian Institute of Science Education and Research - Kolkata, \\
          Mohanpur - 741252, West Bengal, India. }
\email{arghyac@iiserkol.ac.in}

\author{Kirtiman Ghosh}
\affiliation   {Department of Physics and Oklahoma Center for High Energy Physics,
Oklahoma State University, Stillwater, OK 74078-3072, USA. }
\email{kirti.gh@gmail.com}

\author{Sujoy Poddar}
\affiliation {Netaji Nagar Day College, 170/436, N.S.C. Bose Road, Kolkata - 700092, India.}
\email{sujoy.phy@gmail.com}

%\homepage[]{Your web page}
%\thanks{}
%\altaffiliation{}

%Collaboration name if desired (requires use of superscriptaddress
%option in \documentclass). \noaffiliation is required (may also be
%used with the \author command).
%\collaboration can be followed by \email, \homepage, \thanks as well.
%\collaboration{}
%\noaffiliation

\date{\today}
\begin{abstract}
In this work we study the collider phenomenology of a  compressed supersymmetric model with gluino ($\gl$) and the lightest neutralino ($\lspone$). All other 
sparticles are assumed to be heavy. We consider gluino pair production at the 14 TeV LHC and present the mass reach of gluino 
as a function of mass splitting between gluino and the the lightest neutralino. We find that the gluino mass below 1 TeV can be excluded 
at 95$\%$ CL with integrated luminosity of 100 $\ifb$ for extreme degenerate case where mass separation between gluino and 
the lightest neutralino is about 20 GeV. On the other hand, the lower bound on the mass of gluino increases to 1.2 - 1.3 TeV if the mass splitting between 
the  gluino and $\lspone$ is about 200 GeV.  This result shows that for degenerate gluino, the current mass limit may extend approximately 
400-500 GeV at 14 TeV LHC.       
\end{abstract}
%\begin{abstract}
%In this work we consider a  compressed supersymmetric model with a gluino ($\gl$) and lightest neutralino ($\lspone$) and all other 
%sparticles are assumed to be heavy. We consider gluino pair production at the 14 TeV LHC and present the mass reach of gluino 
%as a function of mass difference between gluino and the neutralino. We find that the gluino mass below 1 TeV can be excluded 
%at 95$\%$ CL with integrated luminosity of 100 $\ifb$ for extreme degenerate case where mass separation between gluino and 
%the ($\lspone$) is about 20 GeV,  . However, the lower bound 
%on mass of gluino raises to 1.2 - 1.3 TeV if the mass splitting between the LSP and gluino is about 200 GeV.  This 
%results shows that for degenerate gluino, the current mass limit may extend approximately 400-500 GeV at the 14 TeV LHC.       
%\end{abstract}

% insert suggested PACS numbers in braces on next line
%\pacs{ 1 1 1 1 }
% insert suggested keywords - APS authors don't need to do this
%\keywords{}

%\maketitle must follow title, authors, abstract, \pacs, and \keywords
\maketitle

% body of paper here - Use proper section commands
% References should be done using the \cite, \ref, and \label commands

%% \section{Introduction}
\noindent
Supersymmetry (SUSY) (see e.g., \cite{Martin:1997ns}) is one of the most promising and 
phenomenologically well-studied extensions of the Standard Model (SM). It
can explain many unsolved issues of the SM, e.g., the mass hierarchy problem, the gauge
coupling unification, viable candidate for cold dark matter (DM), 
the origin of  electroweak (EW) symmetry breaking mechanism etc. 
The major goals of the ongoing Large Hadron Collider (LHC) experiment is to unveil the 
mystery of EW symmetry breaking mechanism and to search for SUSY signals. Depending upon the 
SUSY breaking mechanisms, various types of models can be realized in principle.  Constrained MSSM (cMSSM) \cite{Kane:1993td} 
is one of the SUSY models which draws much attention to the particle physics community due to 
small number of parameters which make this model highly predictive. For this reason, two major collaborations of the LHC, 
ATLAS and CMS have searched for the cMSSM from the very beginning of the LHC run in many different final states. 
In the  $R$-parity  conserving model, SUSY particles are produced in pairs and the lightest supersymmetric 
particle (LSP) must be stable. In most of the cases, the lightest neutralino ($\lspone$), being the LSP can be a good candidate for cold DM. 
The generic signature of SUSY search comprises of multi-jets + leptons + large amount of missing transverse energy 
($\met$) which arises due to cascade decays of squarks and gluino into jets, leptons and $\lspone$. Here $\lspone$ is the 
primary source of $\met$ which escapes the detector like neutrinos.

In cMSSM, the gaugino mass universality condition at the high scale leads to the fact that the gaugino masses at 
TeV scale ($m_i$) are proportional to the corresponding gauge coupling constants $\alpha_i$'s: 
$m_3:m_2:m_1 \sim \alpha_3:\alpha_2:\alpha_1\sim 6:2:1$. Therefore it is evident that the gluino is the heaviest gaugino 
and it is almost 6 times heavier than the LSP in that particular model. Jets produced from  the decay of $\gl$, i.e., 
$\tilde{g} \rightarrow q \bar{q} {\tilde \chi_i^0} $ are very energetic resulting in signatures having sufficient 
amount of $\met$ as well as effective mass ($M_{eff}$). Here $M_{eff}$ is defined as the scalar  sum of $P_T$ of jets, 
$P_T$ of leptons (wherever leptons are present) and $\met$. These two kinematic variables ($\met$, $M_{eff}$) can be 
efficiently used to discriminate SUSY signals from the SM backgrounds. The CMS \cite{Chatrchyan:2013lya} and ATLAS {\cite{ATLAS-CONF-2013-047} 
collaborations have searched for SUSY in jets + leptons + $\met$ channel and in the absence of significant excess of signal events over the SM backgrounds, they put stringent bounds on the masses of squarks and the gluino in the framework of cMSSM using 7/8 TeV data.
For example, with integrated luminosity ($\lum$) = 20.3 fb$^{-1}$, equal masses of squarks and gluino are excluded below 
1.7 TeV in cMSSM scenario  with ${\rm tan}\beta=30,~A_0=-2m_0~{\rm and}~\mu>0$ in the  jets + 0$l$ +$\met$ channel from 
8 TeV LHC data \cite{ATLAS-CONF-2013-047}. In a recent Monte Carlo simulation, ATLAS collaboration has shown that equal squarks and gluino masses 
less than around 3 TeV can be excluded at 14 TeV LHC with $\lum$ = 300 $\ifb$ assuming $\mlspone$= 0 GeV 
\cite{ATL-PHYS-PUB-2013-002}. \\

It is to be noted that the hardness of the signal jets/leptons depends only on the relative mass separation 
between squark/gluino  and the LSP, not on the masses of the produced particles. In case of quasi-degenerate mass spectrum, P$_T$ of jets or 
leptons arising from the decay of SUSY particles will be soft and it may even fall below the detector acceptance level. 
In other words, depending on the relative mass difference between squark/gluino and LSP,  the amount of visible (missing) 
transverse energy becomes smaller compared to that in the cMSSM model. In such cases, even if the pair production 
cross-section of squark and/or gluino is large, signal may not be observed over backgrounds because of the poor acceptance.
Consequently the bound on squark/gluino mass will be drastically reduced. Attention has been paid by several authors 
\cite{Alves:2010za, LeCompte:2011cn, LeCompte:2011fh,Alvarez:2012wf,Dreiner:2012gx,Bhattacherjee:2012mz, Dreiner:2012sh} in this direction which basically opens up significant regions of 
parameter space with low  squark/gluino mass. ATLAS collaboration has also searched for compressed SUSY scenarios assuming 
some specific simplified models \cite{Aad:2012fqa, ATLAS-CONF-2013-047}. In particular, if we consider a model with only $\gl$ and $\lspone$, 
$\mgl$ up to 500-550 GeV is excluded  from 7/8 TeV analysis by ATLAS in case of extreme degeneracy $\Delta m (\gl -\lspone)=$10-20 GeV 
which is consistent with the phenomenological results \cite{Alves:2010za, LeCompte:2011cn, LeCompte:2011fh,Dreiner:2012gx,Bhattacherjee:2012mz, 
Dreiner:2012sh}. We can conclude from the above discussion that the limit on $\mgl$ is considerably weaker than that with non-degenerate 
scenarios like cMSSM.

On the other hand, the initial state radiation (ISR) resulting from the quark/gluon legs of squark/gluino production process  depends only on the 
scale of the interaction and the colour structure. Hence, the hardness of ISR jets do not depend on the relative mass 
separation between the produced particles and the corresponding decay products in contrast to jets produced in the decay. 
The mass of the strongly interacting SUSY particles (squarks and gluino) under consideration may be of the order of a few hundreds 
of GeV to TeV. Thus, we expect that the ISR jets arising along with the pair production of squarks/gluino might be hard 
enough to be detected \cite{Alwall:2009zu}. In presence of hard radiation, the pair of squarks/gluino system would recoil 
against the ISR jets giving rise to  $\met$ comparable with the transverse momentum of the jets. Therefore, if squarks/gluino are 
nearly degenerate with the LSP, the jets (coming from ISR) + $\met$ is the only detectable signature. However, in such cases the 
jet multiplicity is smaller than the multiplicity expected in non-degenerate 
SUSY scenarios and thus, the SM backgrounds could be severe. It is not possible to predict the future reach of quasi-degenerate 
SUSY from the study of non-degenerate scenarios. Phenomenologically, it is therefore relevant as well as important to study the future prospect of 
degenerate scenarios at the 14 TeV LHC.

In this work, we are particularly interested in the SUSY scenario with quasi-degenerate $\gl$-$\lspone$.  One may think about the motivation to consider such compressed 
 spectrum in the framework of SUSY.  We already know that there is a plethora of well motivated models which do not obey the cMSSM  mass hierarchy between the gauginos.  
 For example,  (i)  quasi degenerate $\gl- \lspone$ scenario arises in some specific form of GUT model where supersymmetry is broken by gauge mediation \cite{Raby:1997bpa,Mafi:2000kg}.  
 (ii) It it shown in  Ref~\cite{Murayama:2012jh} that compressed SUSY mass spectrum is a automatic outcome of supersymmetry breaking via boundary conditions in compact extra dimensions,   
 (iii) In the pure gravity mediation model, the gaugino mass relation can be significantly modified in the presence of axion and in some part of the parameter space, gluino can be  nearly degenerate with the $\lspone$ \cite{Nakayama:2013uta}.   
(iv).  Models with a small mass difference between gluino and LSP may be favoured by precision gauge coupling unification \cite{Krippendorf:2013dqa}.  
 Also quasi-degenerate gluino-neutralino scenario is well motivated from dark matter point of view. It is well known that the annihilation of bino like LSP gives rise to the relic density which is too large compared to the DM relic density data measured  by PLANCK \cite{Ade:2013zuv}. Observed relic density can be explained if bino like $\lspone$ coannihilates with nearly degenerate gluino \cite{Profumo:2004wk}.  Apart from this, light gluino is always favoured from the consideration of naturalness. 

In our analysis  we consider $\gl$ pair production at the 14 TeV LHC with subsequent decay of $\gl$ into 
light quarks and $\lspone$. We have focused on 2/3/4 jets + $\met$ signatures\footnote{ Mono-jet plus $\met$ is also a viable signal of degenerate gluino production. However, 7/8 TeV results \cite{Dreiner:2012gx,Bhattacherjee:2012mz} show that it is weaker than the conventional 2/3 jets + $\met$ searches.} which 
are mainly important for compressed SUSY models. As mentioned before, for compressed spectrum, 
jets mainly come from ISR. If we just generate parton level $\gl$ pair events and pass these events through event generator for 
hadronization, the ISR jets  will be produced by the event generator at the time of showering. However, the ISR jet spectrum 
crucially depends on the factorization scale. Depending upon the choice of factorization scale, the spectrum might be harder 
as well as softer. Consequently it induces large amount of systematic uncertainty in the calculation of mass limits. Monte carlo events 
generators like PYTHIA or HERWIG use different parton shower algorithms. Depending upon the choice of  generators we may get somewhat 
different results.  In order to avoid this, 
we have to generate $\gl$ pair in association with additional jets at the matrix element level. However, in order to get rid 
of the double counting problem at the time of parton showering we must adopt a particular matching prescription. 
This also reduces the systematic uncertainty coming from the different parton shower schemes although monte carlo 
event generators as a whole introduces a large amount of systematic uncertainty which can not removed by matching 
prescription. \\

SUSY mass spectra and decay processes are generated by SUSY-HIT \cite{Djouadi:2006bz}. Throughout this work 
LSP is assumed to be bino dominated and all SUSY particles except $\gl$ and $\lspone$ are set beyond the 
reach of 14 TeV LHC. The DM relic density is computed by using MicrOMEGAs \cite{Belanger:2001fz} for our scenario. We have used 
MadGraph 5 \cite{Alwall:2011uj} for generating both signal events and SM backgrounds at the parton level. For signal events 
we have generated $\gl \gl$ and $\gl \gl$ plus one additional jet at the parton level. MadGraph can calculate 
$\gl$ pair up to 2 extra jets at the matrix element level. However, for scanning purpose, it is not so efficient because
it is very much time consuming and we have checked that the inclusion of the second additional jet does not change the 
result appreciably. Dominant backgrounds for our analysis are $Z$ + jets, $W$ + jets and $t \bar t$ + jets\footnote{QCD multijets, 
electroweak gauge boson pairs may be considered as potential backgrounds. However, after inclusion 
of all cuts discussed in Table~\ref{tab1}, these backgrounds turn out to be negligibly small and hence they are not included in 
our analysis.}. $W/Z$ backgrounds are generated up to additional 4 jets and for  $t \bar t$ events we have considered up 
to 2 jets. Subsequently signal and background events are passed through PYTHIA \cite{Sjostrand:2006za} for showering, decay, 
hadronization etc. For matching purpose we have used MLM \cite{Mangano:2006rw} prescription as implemented in MadGraph 5 for both signal and 
backgrounds. Finally the events are fed to fast detector simulator package Delphes \cite{Ovyn:2009tx} for object reconstructions\footnote{Jets are 
reconstructed with  anti-$k_t$ algorithm with $R = 0.4$.}. For signal, NLO cross sections are obtained 
from  PROSPINO 2.1 \cite{Beenakker:1996ed}. For $t \bar t$ background we have used NLO cross section \cite{ttbarnlo}. 
The K factors for electroweak (EW) processes ($W/Z$ backgrounds) are generally small $\sim$ 1.15 -1.20 \cite{Catani:2009sm}. 
We do not explicitly compute the K factors for those processes, rather we conservatively assume K = 1.2 for $W/Z$ productions.

The important variables used for SUSY search are $\met$ and $M_{eff}$ which are described previously. It is thus important to study 
the shape of these distributions for degenerate SUSY and compare it with SM backgrounds. In Fig.~\ref{dist} we have illustrated the 
normalised $M_{eff}$(incl.) distribution after some nominal cuts where $M_{eff}(incl.)$ is defined as the scalar sum of 
$\met$ and $P_T$ of all jets satisfying $P_T >$ 40 GeV. For illustration, we choose two benchmark points, one with large 
$\Delta m$ ($\mgl$ = 800 GeV, $\mlspone$ = 100 GeV) and another with small mass splitting ($\mgl$ = 800 GeV, $\mlspone$ = 780 GeV). 
Fig.~\ref{dist} shows a clear distinction between signal and background distributions for non degenerate scenario as expected. 
For degenerate benchmark point, $M_{eff}$(incl.) peaks at lower value similar to SM backgrounds. However, for 
higher values of $M_{eff}$(incl.) SM backgrounds fall rapidly in comparison to both SUSY benchmark points and a 
strong cut on $M_{eff}$(incl.) (e.g., $>$ 1.5 TeV) may be used to probe degenerate SUSY. It is also clear from Fig.~\ref{dist}  
that the above mentioned cut is more effective for non degenerate spectrum. Similarly, $\met$ distribution has also the same 
behaviour. We should note that the discovery of SUSY signal depends on the tail of the background distributions and 
care should be taken to generate  the tail which consists of very low probability events. To verify the stability of our result we have 
also generated sufficient number of events ($\sim$ 100 $\ifb$) by applying $M_{eff}$(incl.) cut at the parton level using MadGraph.  
We have checked that the backgrounds estimated using these two methods (with and without parton level $M_{eff}$(incl.) cut) 
are quite consistent with each other.        

%----------------------------------------------------------------
\begin{figure}[t]
\begin{center}
\includegraphics[angle =360, width=0.8\textwidth]{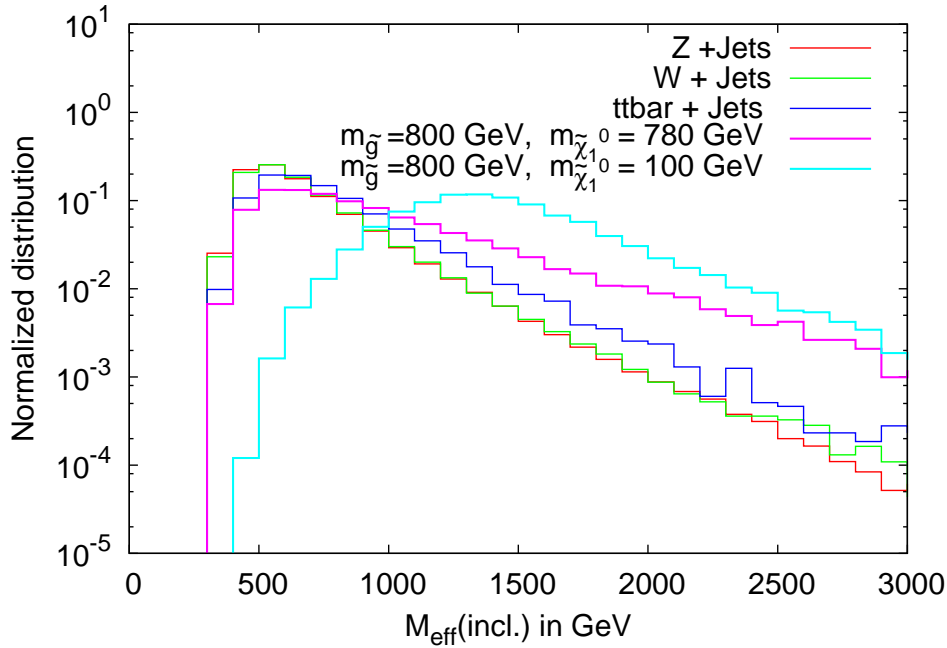}
\end{center}
\caption{ {\footnotesize  Normalized $M_{eff}$(incl.) distribution for different SM backgrounds and SUSY benchmark points after 
a few nominal cuts (lepton veto, 
$P_T(j_1)>$130 GeV, $P_T(j_2)>$60 GeV and $\met >$160 GeV). }}\label{dist}
\end{figure}
%--------------------------------------------------

Motivated by the above discussion we study the prospect of degenerate $\gl$ search at the early run of 14 TeV LHC.  
We vary the mass difference between $\gl$ and $\lspone$ from 20 to 200 GeV for $\mgl$ in the range 550 GeV to 1.5 TeV and 
by changing $\Delta m= (\mgl -\mlspone)$, we calculate the expected future limit in the $\mgl$ - $\Delta m$ plane. For analysis we choose three different strategies described below.

{\bf Strategy A : } The ATLAS collaboration has studied the prospect of SUSY search in 2-4 jets + $\met$ channel 
at the 14 TeV run \cite{Aad:2009wy}. We closely follow their analysis for our study. Moreover 
we also vary $M_{eff}$ cuts, which is not considered by ATLAS collaboration. 
 
{\bf Srategy B : } We adopt the cuts used in latest ATLAS 8 TeV analysis in the jets +0l\footnote
{Events with isolated electron/muon with P$_T>$ 10 GeV are rejected.} + $\met$ search channels with 
$\lum = $20.3 $\ifb$ \cite{ATLAS-CONF-2013-047}. As our signal consists of low jet multiplicities 
(mainly ISR jets), we do not consider signal regions with jet multiplicity greater than 4. Details of the cuts 
used in our analysis is presented in Table~\ref{tab1}.

It is found that better results are obtained using Strategy B compared to Strategy A. For this reason we do not further 
discuss about the ATLAS old analysis (Strategy A). 

{\bf Srategy C : }  Since Strategy B is optimized for 8 TeV analysis, we expect more stronger cuts on variables for 
14 TeV. Therefore, in order to enhance the significance, we try to optimize different variables like $P_T$ of 
jets, $\met$, $M_{eff}$(incl.), $\met$/ $M_{eff}(N_j)$ etc. for our 14 TeV analysis. We define more than 200 different 
possible combination of cuts and we choose 4 cut sets which give best results for degenerate scenario as described 
in Table~\ref{tab1}. In Table~\ref{tab1} SRA-OT stands for optimized set of cuts for 2-jet signal region with T 
denoting tight cut on  $M_{eff}$(incl.). Other notations have similar meaning, defined in Table~\ref{tab1}. 

%======================================================================================
\begin{table*}[!htb]
\begin{center}\
\tiny  {
\footnotesize {
\begin{tabular}{|c||c|c|c|c|c|c|c|}
\hline
  	   &   \multicolumn{7}{c|}{Channel}\\
\cline{2-8}
     Cuts              		&SRA-OT	&SRB-M	&SRB-T	&SRB-OT	&SRC-T	&SRC-OM	&SRC-OT	\\
                   		& (2j) 	&(3j)	&(3j)	&(3j)	&(4j)	&(4j)	&(6j)	\\
       \hline

$\etslash$ [GeV] $>$		&200	&160	&160	&200	&160	&200	&200	\\ 
\hline
$P_T(j_1)$ [GeV] $>$    	&200	&130	&130	&150	&130	&150	&150	\\ 
\hline
$P_T(j_2)$ [GeV] $>$    	&100	&60	&60	&80	&60	&80	&80	\\ 
       \hline
$P_T(j_3)$ [GeV] $>$		&-	&60	&60	&80	&60	&80	&80	\\ 
\hline
$P_T(j_4)$ [GeV] $>$		&-	&-	&-	&-	&60	&80	&80	\\ 
\hline
$\del \phi(jet_i,\etslash)_{min}$ &0.4	&0.4	&0.4	&0.4	 			&\multicolumn{3}{c|}{$0.4(i=[1,2,3])$ }\\
				  &(i=1,2)**	&(i=1,2,3)&(i=1,2,3)&(i=1,2,3) 	&\multicolumn{3}{c|}{0.2($P_T>$40 GeV jets)}\\
%\hline
% $\etslash / H_T$ $>$ &-	&	&	&	&	&	&	\\ 
\hline
$\met$/ $M_{eff}(N_j)$$>$	&0.4	&0.3	&0.4	&0.4	&0.25	&0.4	&0.3	\\ 
\hline 
$M_{eff}$(incl.)[GeV] $>$	&2400	&1800	&2200	&2400	&2200	&2200	&2400	\\ 
\hline
       \hline

$Z+jets$[fb]			&4.0	&18.4	&3.9	&1.8	&2.8	&0.66	&0.9	\\ 
\hline
$W+jets$[fb]		   	&1.15	&8.6	&1.2	&0.6	&1.2	&0.15	&0.4	\\ 
\hline
$t\bar t +jets$	[fb]		&0.25	&4.3	&0.3	&0.1	&0.4	&0.12	&0.2	\\ 
       \hline
       \hline
Total SM background[fb]		&5.4	&31.3	&5.4	&2.5	&4.4	&0.93	&1.5	\\ 
\hline
\hline
%$<\epsilon \sigma>_{obs}^{95}$[fb]&	&	&	&	&	&	&	\\ 
%\hline 
Upper limit on $N_{BSM}$ at 95$\%$ CL	  &59	&381	&68	&35	&57	&17	&23	\\ 
Sys. Un.=20 $\%$, $\lum$=30$\ifb$ &	&	&	&	&	&	&	\\ 
\hline 
Upper limit on $N_{BSM}$ at 95$\%$ CL	   &116	&636	&116	&59	&97	&27	&39	\\ 
Sys. Un.=10 $\%$, $\lum$=100$\ifb$ &	&	&	&	&	&	&	\\ 
\hline 
\hline
       \end{tabular}
}
}
       \end{center}
          \caption{ {\footnotesize Selection criteria considered in our analysis at the 14 TeV LHC and upper limit on the number of 
          BSM events. SRB-M, SRB-T, SRC-T signal regions are taken from \cite{ATLAS-CONF-2013-047} and signal 
          regions are denoted by the same convention as ATLAS. SRA-OT, SRB-OT, SRC-OM, SRC-OT signal regions 
           are obtained by optimization of cuts. O,T,M denote optimized, tight and medium respectively. Leptons are 
           vetoed in all of the signal regions. $M_{eff}(N_j)$ is constructed from only the leading N jets(indicated in the parenthesis of 2nd row.  
           and $\met$ . ** For SRA-OT, $\del \phi(jet_i,\etslash)_{min}$  cut is applied for 
           $i$ = 1 to 3 if $P_T(j_3) > $40 GeV. }}
\label{tab1}
          \end{table*}

%======================================================================================
Seven signal regions which are relevant for the search of degenerate gluino are described in Table~\ref{tab1}. 
Contributions of individual SM backgrounds after the final cut are also shown in Table~\ref{tab1}. 
We can see from Table~\ref{tab1} that the dominant backgrounds are arising from $Z(W)$ + jets production followed 
by $Z \rightarrow \nu \nu$ and $W \ra l \nu$. The first one is the irreducible background for jets + $0l$ + 
$\met$ signature whereas the second one contributes only when the lepton is not reconstructed. 

In order to estimate the exclusion limit on $\mgl$ we need to know the systematic uncertainties. 
We expect that for early run of 14 TeV LHC systematic uncertainties will be large and it may be reduced with 
 the collection of more data. We have computed the upper limit on the number of SUSY events ($N_{BSM}$) for integrated 
$\lum$ = 30 and 100 $\ifb$ assuming systematic uncertainties to be 20\% and 10\% respectively using 
Bayesian method at 95 $\%$ confidence level (CL). The last two rows of Table~\ref{tab1} represent these numbers.

%----------------------------------------------------------------
\begin{figure*}[tb]
\begin{center}
\includegraphics[angle = 270, width=0.6\textwidth]{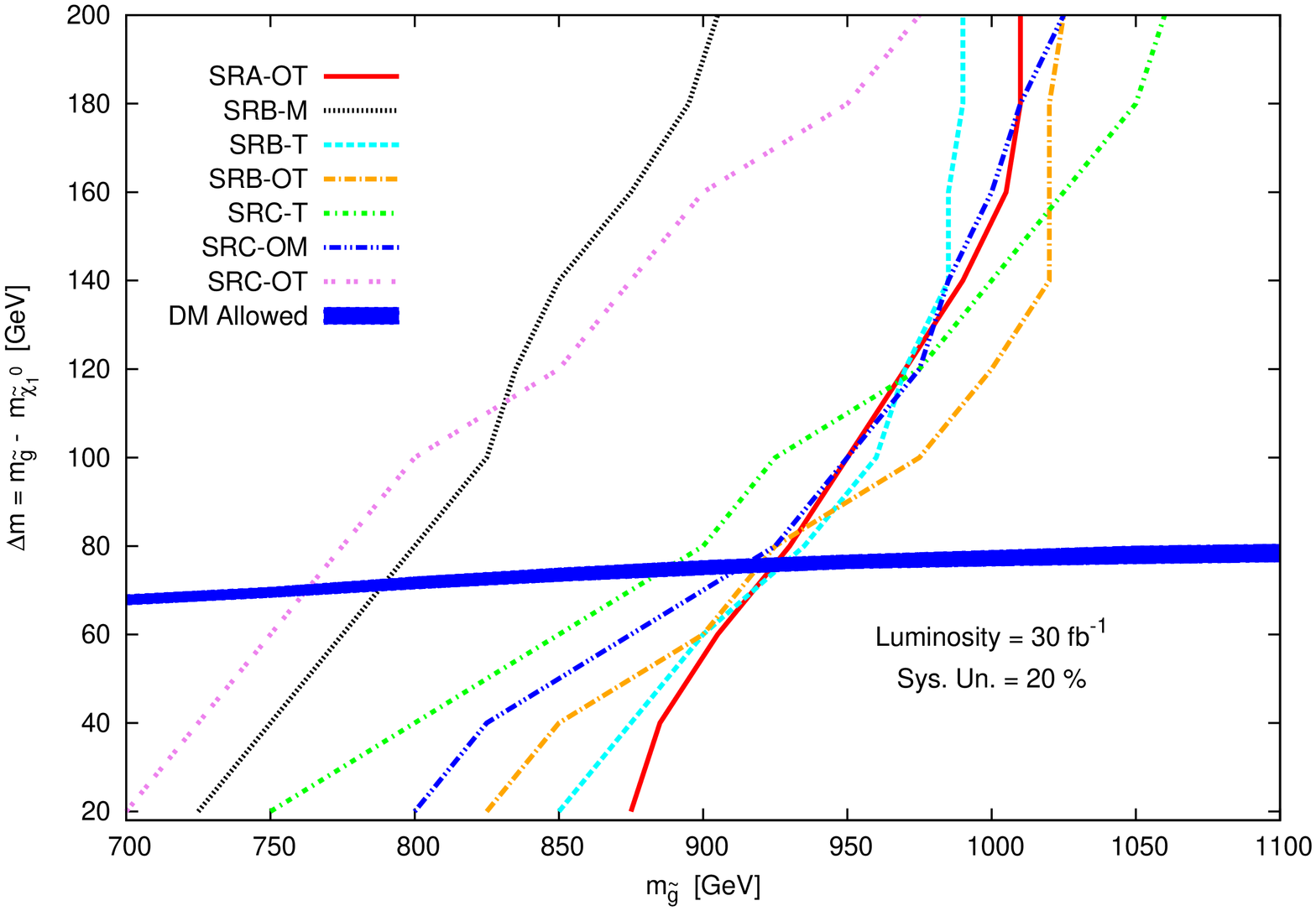}
\hspace{-1.4cm}
\includegraphics[angle = 270, width=0.6\textwidth]{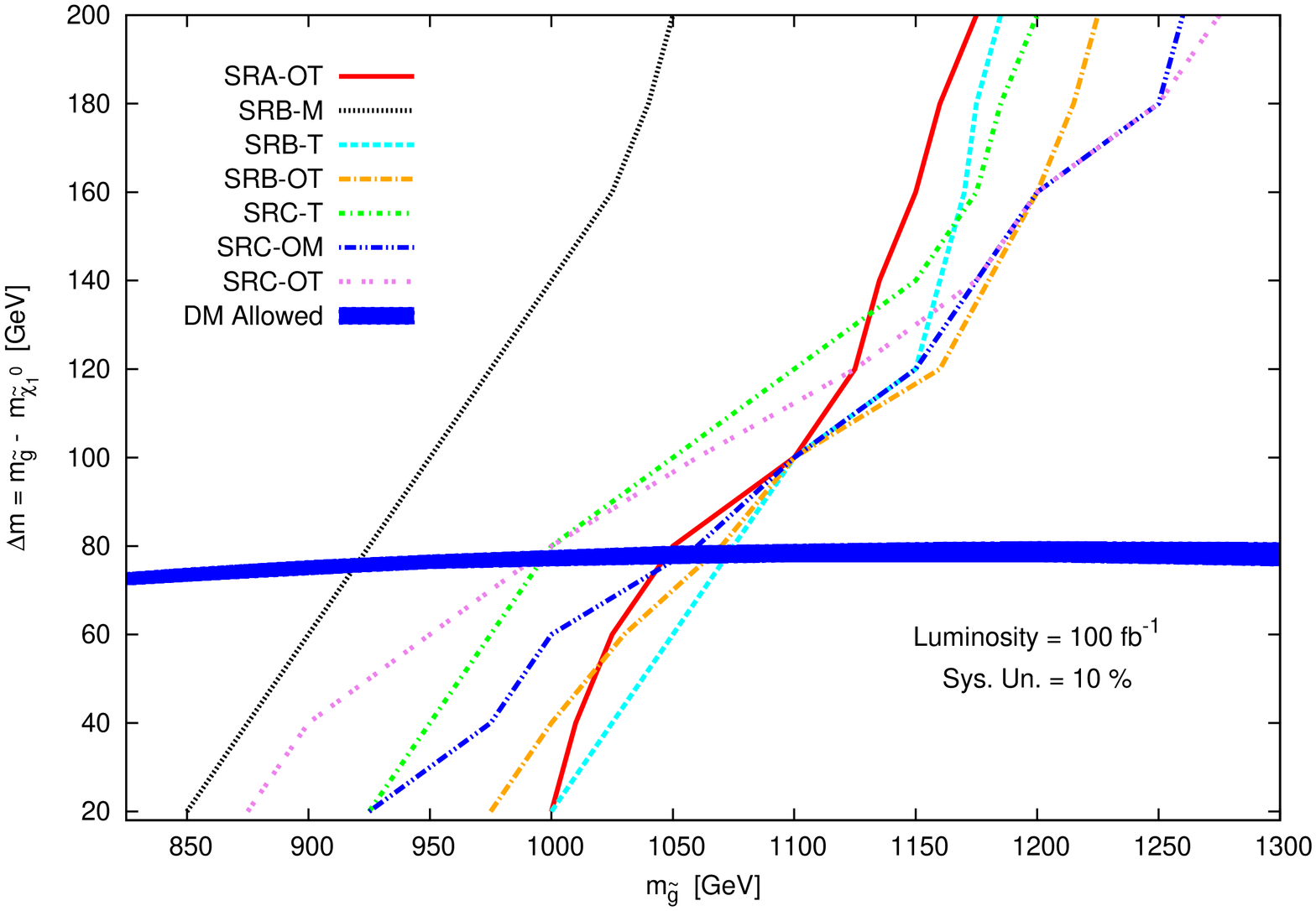}
\end{center}
\vspace{-0.5cm}

\caption{ {\footnotesize 95 $\%$ CL exclusion limit in the $\mgl$  vs $\Delta m = \mgl - \mlspone$ plane 
with integrated $\lum$ = 30 (top) and 100 (bottom $\ifb$ assuming systematic uncertainty to be 20\% and 10 $\%$ respectively. 
Exclusion lines correspond to seven signal regions as described in  Table~\ref{tab1}. Shaded (blue) region represents  3 $\sigma$ 
band of the allowed DM relic density measured by PLANCK collaboration.  } }\label{bound}
\end{figure*}
%--------------------------------------------------

In Fig.~\ref{bound} we present the exclusion limits on gluino mass in the $\mgl$  vs $\mgl - \mlspone$ plane using seven signal regions 
defined in Table~\ref{tab1} with $\lum =$ 30 $\ifb$ and 
systematic uncertainty = 20 $\%$, SRA-OT signal region gives the best limit: $\mgl >$ 875 GeV for extreme degenerate 
scenario ($\Delta$m = 20 GeV) and for $\Delta$m = 200 GeV we obtain $\mgl$ $>$ 1050 GeV using signal region SRC-T. Again with 
increased $\lum =$ 100 $\ifb$ and less systematic uncertainties (10 $\%$) SRB-T or SRB-OT gives the best exclusion limit 
on $\mgl$ ($>$ 1 TeV) for $\Delta$m = 20 GeV. For $\Delta$m = 200 GeV SRC-OT excludes gluino mass below 1275 GeV. \\

Here we remind that the mass limit depends significantly on the systematic uncertainty and  the bounds presented in 
Fig.~\ref{bound}, assuming 10\% uncertainty for  $\lum =$ 100 $\ifb$ may be optimistic. It is a very challenging task 
to reduce the systematic uncertainty to such a small value in future LHC run and it is thus important to study the 
sensitivity of limits on systematic uncertainty. If we consider high luminosity ( $\lum$= 300 $\ifb$) option with 20\% 
systematic uncertainty, gluino mass limit is expected to be about 925 GeV, whereas this limit is reduced to about 800 GeV 
for 30\% systematic uncertainty. Assuming systematic uncertainty = 20\% and $\lum$ = 300 $\ifb$, 
the 5$\sigma$ discovery reach of gluino is just 725 GeV for extreme degeneracy, which is close to current LHC exclusion limit.    
Although, our main focus is to search for degenerate gluino, the results, presented in this paper can be directly used 
for degenerate squark production, dark matter search and other degenerate new physics scenarios like minimal version of the 
Universal Extra Dimension model \cite{Appelquist:2000nn}.  

%%%%%%%%%%%%%%%%%%%%%%%%%%%%%%%%%%%%%%%%%%%%%%%%%%%%%%%%%%%%%%%%%%%%%%%%%%%%%%%%%%%%%%%%%%%%5
%\begin{table*}[!htb]
%\begin{center}\
%\begin{tabular}{||c||c|c||c|c||}
%\hline
%Signal Region	& \multicolumn{2}{c|}{$\lum = 30\ifb, SU =20\%$} & \multicolumn{2}{c|}{$\lum = 100\ifb, SU =10\%$}	\\
%\cline{2-5}
%		&$\Delta$m = 20 GeV	&$\Delta$m = 200 GeV	 &$\Delta$m = 20 GeV	&$\Delta$m = 200 GeV		\\
%\hline
%SRA30	 	&875		& 1010		&1000		&1175	\\
%\hline
%SRB-M	 	&725		& 905		&850		&1050	\\
%\hline
%SRB-T	 	&850		&990 		&1000		&1185	\\
%\hline
%SRB15	 	&825		&1025		&975		&1225	\\
%\hline
%SRC-T	 	&750		&1060 		&900		&1200	\\
%\hline
%SRC14	 	&800		& 1000		&925		&1250	\\
%\hline
%SRC10	 	&700		& 950		&900		&1275	\\
%\hline
%       \end{tabular}\
%       \end{center}
%           \caption{Limits on $\mgl$. }
%\label{tab2}
%          \end{table*}
%%%%%%%%%%%%%%%%%%%%%%%%%%%%%%%%%%%%%%%%%%%%%%%%%%%%%%%%%%%%%%%%%%%%%%%%%%%%%%%%%%%%%%%%%%%%5

{ \bf \large Conclusion :} 
In this paper we have considered a scenario with gluino as the next to lightest supersymmetric particle (NLSP) 
and the mass difference between gluino and $\lspone$  is smaller compared to the conventional cMSSM 
model. This type of scenario can be realised in various SUSY breaking mechanisms and  $\gl$- $\lspone$ coannihilation 
can explain the observed DM relic density of the universe. Moreover, the present bound on $\mgl$, obtained  from 
LHC 7/8 TeV data is rather weaker ($\mgl >$ 500-550 GeV).  It is thus very important to study the future search prospect of 
such scenario as there is no detailed phenomenological work in this direction for 14 TeV LHC till date. The signal consists 
of small number of jets (dominantly ISR jets), moderate amount of $\met$ and $M_{eff}$(incl.) making the signal challenging 
to discover over huge SM backgrounds. We have investigated the discovery reach  of jets + 0l + $\met$  channels for the 
quasi-degenerate gluino NLSP and neutralino LSP scenario at the 14 TeV LHC with integrated luminosity up to 100 $\ifb$  
using optimized cuts (presented in Table~\ref{tab1}). We have found that it is possible to exclude gluino mass up to 1 TeV 
 for extreme degenerate case at 95 $\%$ CL. For moderately degenerate case ($\Delta$m = 200 GeV), the exclusion limit 
 on $\mgl$ may reach up to (1.2 - 1.3) TeV  at 95 $\%$ CL in the near future. Although we have investigated the degenerate 
 gluino case, this analysis is more generic in nature and it can be applied to any other new physics search using jets (+ 0l) plus 
 small or moderate amount of  $\met$ in the final state.

{ \bf Acknowledgments : }
%\begin{acknowledgments}
The work of BB is supported by World Premier International Research Center Initiative (WPI Initiative), MEXT, Japan.
The work of KG is supported by US Department of Energy, Grant Number DE-FG02-04ER41306.
%\end{acknowledgments}
%\bibliography{paperv3}
%merlin.mbs apsrev4-1.bst 2010-07-25 4.21a (PWD, AO, DPC) hacked
%Control: key (0)
%Control: author (8) initials jnrlst
%Control: editor formatted (1) identically to author
%Control: production of article title (-1) disabled
%Control: page (0) single
%Control: year (1) truncated
%Control: production of eprint (0) enabled
%

\end{document}